\documentclass[twocolumn, pra,showpacs,superscriptaddress]{revtex4}
\usepackage{amssymb}
\usepackage{graphicx}
\usepackage{dcolumn}
\usepackage{bm}
\usepackage{amsmath}

\begin{document}

\title{Analytical solutions for the Rabi model}
\author{Lixian Yu}
\affiliation{School of Physical Science and Technology, Soochow University, Suzhou,
Jiangsu 215006, P. R. China}
\affiliation{Department of Physics, Shaoxing University, Shaoxing 312000, P. R. China}
\affiliation{State Key Laboratory of Quantum Optics and Quantum Optics Devices, Laser
spectroscopy Laboratory, Shanxi University, Taiyuan 030006, P. R. China}
\author{Shiqun Zhu}
\affiliation{School of Physical Science and Technology, Soochow University, Suzhou,
Jiangsu 215006, P. R. China}
\author{Qifeng Liang}
\affiliation{Department of Physics, Shaoxing University, Shaoxing 312000, P. R. China}
\author{Gang Chen}
\thanks{Corresponding author: chengang971@163.com}
\affiliation{State Key Laboratory of Quantum Optics and Quantum Optics Devices, Laser
spectroscopy Laboratory, Shanxi University, Taiyuan 030006, P. R. China}
\author{Suotang Jia}
\affiliation{State Key Laboratory of Quantum Optics and Quantum Optics Devices, Laser
spectroscopy Laboratory, Shanxi University, Taiyuan 030006, P. R. China}

\begin{abstract}
The Rabi model that describes the fundamental interaction between a
two-level system with a quantized harmonic oscillator is one of the
simplest and most ubiquitous models in modern physics. However, this
model has not been solved exactly because it is hard to find a
second conserved quantity besides the energy. Here we present a
unitary transformation to map this unsolvable Rabi model into a
solvable Jaynes-Cummings-like model by choosing a proper variation
parameter. As a result, the analytical energy spectrums and
wavefunctions including both the ground and the excited states can
be obtained easily. Moreover, these explicit results agree well with
the direct numerical simulations in a wide range of the experimental
parameters. In addition, based on our obtained energy spectrums, the
recent experimental observation of Bloch-Siegert in the circuit
quantum electrodynamics with the ultrastrong coupling can be
explained perfectly. Our results have the potential application in
the solid-state quantum information processing.
\end{abstract}

\maketitle

\section{Introduction}

The Rabi model, which was introduced over 70 years ago, describes the
important interaction between a two-level system and a quantized harmonic
oscillator (bosonic mode) \cite{Rabi}. The corresponding Hamiltonian of this
model is written as
\begin{equation}
H_{\text{R}}=\omega a^{\dagger }a+\frac{1}{2}\Omega \sigma _{z}+g\sigma
_{x}(a^{\dagger }+a),  \label{RM}
\end{equation}%
where $a^{\dagger }$ and $a$\ are creation and annihilation operators for
the quantized harmonic oscillator with frequency $\omega $, $\sigma _{i}$ ($%
i=x,y,z$) are the Pauli spin operators, $\Omega $ is the resonant frequency
between two levels, and $g$ is the interaction strength. In modern physics
ranging from quantum optics \cite{Scully}, condensed-matter physics \cite%
{Holstein} to quantum information \cite{Raimond}, the Rabi model has been
one of the simplest and most ubiquitous models. Despite its old age and
central importance, the Rabi model has not been solved exactly because it is
hard to find a second conserved quantity besides the energy. The well-known
method to obtain the energy spectrums and the wavefunctions for the Rabi
model is the numerical diagonalization in a truncated finite-dimensional
Hilbert space \cite{CQH}. However, these numerical results are difficult to
extract the fundamental physics of the Rabi model \cite%
{Crisp,Lamata,Zhu,Gerritsma,Larson,Larson1,Larson2} and to precisely control
the experimental parameters to process quantum information \cite{You}.

To overcome the problem in the analytical considerations, the rotating-wave
approximation, which is valid in the regime $g\ll \omega $, was proposed to
rewrite Hamiltonian (\ref{RM}) as \cite{Jaynes}
\begin{equation}
H_{\text{JC}}=\omega a^{\dagger }a+\frac{1}{2}\Omega \sigma _{z}+g(\sigma
_{-}a^{\dagger }+\sigma _{+}a).  \label{JC}
\end{equation}%
Interestingly, the Jaynes-Cummings model with $U(1)$ symmetry has a
conserved quantity $C=a^{\dagger }a+\frac{1}{2}(\sigma _{z}+1)$, and thus,
it can be solved easily in the subspaces $\{\left\vert \uparrow
,n\right\rangle ,\left\vert \downarrow ,n+1\right\rangle \}$. Meanwhile, the
Jaynes-Cummings model can successfully describe quantum dynamics of optical
cavity electrodynamics with strong coupling. However, in the recent
investigations about the solid-state quantum electrodynamics with the
ultrastrong coupling ($g\sim 0.1\omega $) \cite%
{Gunter,Anappara,Todorov,Fedorov,Hofheinz,LaHaye,Niemczyk,Peropadre,Geiser,Albert}%
, the rotating-wave approximation breaks down and the system's dynamics must
be governed by the Rabi model. Irish put forward a generalized rotating-wave
approximation to solve the Rabi model. This method can be used to derive the
analytical energy spectrums of Hamiltonian (\ref{RM}) in the ultrastrong
coupling successfully \cite{Irish}. However, this method works reasonable
only for the negative detuning $(\Omega <\omega )$ \cite%
{Liu,Hausinger,Albert1}. Recently, Braak used the property of Z$_{2}$
symmetry of the Rabi model to obtain its analytical solutions, which are,
however, dependent of the composite transcendental function defining through
its power series in the interacting strength $g$ and the frequency $\omega $
\cite{Braak}. In our previous work, we put forward a generalized variational
method to analytically obtain the ground-state energy of the Rabi model,
which agrees well with the numerical simulation in all regions of the
resonant frequency of the two-level system including the negative detuning $%
(\Omega <\omega )$, the resonant case $(\Omega =\omega )$, and especially
the positive detuning $(\Omega >\omega )$. Unfortunately, our introduced
method cannot be used to consider the excited-state energy spectrum \cite%
{Zhang}.

In this paper, we present a simple and straightforward method to
solve the Rabi model in both ground and excited states. We first map
the unsolvable Rabi model (\ref{RM}) into a solvable
Jaynes-Cummings-like model by choosing a proper variation parameter,
as shown in Section II. Thus, the analytical energy spectrums
including the ground and the excited states are obtained in section
III. These derived analytical results agree well with the direct
numerical simulations in a wide range of the experimental
parameters. Moreover, the recent experimental observation of
Bloch-Siegert \cite%
{Fedorov}, which is just the energy shift of the level transition,
in the circuit quantum electrodynamics with the ultrastrong coupling
can be well explained. In section IV, the analytical wavefunctions
and the corresponding experimentally-measurable physics quantities
such as the mean photon number are derived. Also, the obtained mean
photon number can agree well with the numerical simulation. Finally,
some conclusions are remarked in Section V.

\section{Mapping into the Jaynes-Cummings-like model}

When performing a rotation around the $y$ axis, the Rabi model becomes
\begin{equation}
H_{\text{R}}=\omega a^{\dag }a+\frac{\Omega }{2}\sigma _{x}-g(a^{\dag
}+a)\sigma _{z}.  \label{HF}
\end{equation}%
Under a unitary transformation
\begin{equation}
U=\exp [\lambda \sigma _{z}(a^{\dag }-a)]  \label{UT}
\end{equation}%
with $\lambda $ being the dimensionless parameter determined by the
following calculations, an effective Hamiltonian is given by $H_{\text{E}%
}=UHU^{\dagger }$, namely,
\begin{equation}
H_{\text{E}}=H_{1}+H_{2}+H_{3},  \label{EH}
\end{equation}%
where
\begin{equation}
H_{1}=\omega a^{\dag }a-\lambda \omega \sigma _{z}(a^{\dag }+a)+\omega
\lambda ^{2},  \label{H1}
\end{equation}%
\begin{equation}
H_{2}=-g[\sigma _{z}(a^{\dag }+a)-2\lambda ],  \label{H2}
\end{equation}%
\begin{equation}
H_{3}=\frac{\Omega }{2}\{\sigma _{x}\cosh [2\lambda (a^{\dag }-a)]+i\sigma
_{y}\sinh [2\lambda (a^{\dag }-a)]\}.  \label{H3}
\end{equation}

Since $\cosh (y)$ and $\sinh (y)$ are the even and odd functions
respectively, the terms $\cosh [2\lambda (a^{\dag }-a)]$ and $\sinh
[2\lambda (a^{\dag }-a)]$ can be expanded as
\begin{eqnarray}
\cosh [2\lambda (a^{\dag }\!-\!a)]\!\!\! &=&\!\!\!G_{0}(N)+G_{1}(N)a^{\dag
2}+a^{2}G_{1}(N)+\ldots ,  \label{Cosh} \\
\sinh [2\lambda (a^{\dag }\!-\!a)]\!\!\! &=&\!\!\!F_{1}(N)a^{\dag
}-\!a\!F_{1}(N)+  \notag \\
&&\!F_{2}(N)a^{\dag 2}\!-\!a^{2}F_{2}(N)+\!\ldots ,  \label{Sinh}
\end{eqnarray}%
where $G_{i}(N)$ ($i=0,1,2,\cdots $) and $F_{j}(N)(j=1,2,\cdots )$ with $%
N=a^{\dag }a$ are the coefficients that depend on the dimensionless
parameter $\lambda $ and the photon number $n$. In general, the multi-photon
process is weak in the Rabi model \cite{Scully}. It means that the terms of
the high-order terms for $a$ and $a^{\dag }$ in Eqs. (\ref{Cosh}) and (\ref%
{Sinh}) can be eliminated. As a result, the effective Hamiltonian (\ref{EH})
reduces to the form
\begin{eqnarray}
H_{\text{E}} &=&\omega a^{\dag }a-(\lambda \omega +g)\sigma _{z}(a^{\dag
}+a)+\omega \lambda ^{2}+2\lambda g+  \notag \\
&&\frac{1}{2}\Omega \{\sigma _{x}G_{0}(N)+i\sigma _{y}[F_{1}(N)a^{\dag
}-aF_{1}(N)]\}.  \label{MAT}
\end{eqnarray}

In the eigenstates of $\sigma _{x}$ with $\sigma _{x}\left\vert \pm
x\right\rangle =\pm \left\vert \pm x\right\rangle $, the Pauli spin
operators become $\sigma _{z}\rightarrow -(\tau _{+}+\tau _{-})$ and $\sigma
_{y}\rightarrow -i(\tau _{+}-\tau _{-})$, where $\tau _{z}=\left\vert
+x\right\rangle \left\langle +x\right\vert -\left\vert -x\right\rangle
\left\langle -x\right\vert $, $\tau _{+}=\left\vert +x\right\rangle
\left\langle -x\right\vert $ and $\tau _{-}=\left\vert -x\right\rangle
\left\langle +x\right\vert $, and the effective Hamiltonian (\ref{MAT})
turns into
\begin{eqnarray}
H_{\text{E}} &=&\omega a^{\dag }a+\omega \lambda ^{2}+2\lambda g+\frac{1}{2}%
\Omega G_{0}(n)\tau _{z}+  \notag \\
&&R_{r}(\tau _{+}a+\tau _{-}a^{\dag })+R_{ar}(\tau _{+}a^{\dag }+\tau _{-}a),
\label{ESE}
\end{eqnarray}%
where
\begin{equation}
G_{0}(n)=\left\langle n|\cosh [2\lambda (a^{\dag }-a)]|n\right\rangle ,
\label{GN}
\end{equation}%
\begin{equation}
R_{r}=\lambda \omega +g-\frac{1}{2}\Omega f_{1}(m,n),  \label{RWC}
\end{equation}%
\begin{equation}
R_{ar}=\lambda \omega +g+\frac{1}{2}\Omega f_{1}(m,n)  \label{ARWC}
\end{equation}%
with $f_{1}(m,n)=\left\langle m\right\vert F_{1}(N)a^{\dag }\left\vert
n\right\rangle /\sqrt{n+1}=\left\langle n+1\right\vert F_{1}(N)a^{\dag
}\left\vert n\right\rangle /\sqrt{n+1}$. It is straightforward to calculate
that $G_{0}(n)=\left\langle n|e^{2\lambda (a^{\dag }-a)}+e^{-2\lambda
(a^{\dag }-a)}|n\right\rangle /2=\left\langle n|e^{2\lambda (a^{\dag
}-a)}]|n\right\rangle =e^{-2\lambda ^{2}}L_{n}(4\lambda ^{2})$ and $%
f_{1}(m,n)=f_{1}(n+1,n)=$ $\left\langle n+1|\sinh [2\lambda (a^{\dag
}-a)|n\right\rangle =2\lambda e^{-2\lambda ^{2}}L_{n}^{1}(4\lambda
^{2})/(n+1)$, where $L_{n}(y)$ is the Laguerre polynomial and $L_{n}^{1}(y)$
is the associated Laguerre polynomial. Thus, if the dimensionless parameter $%
\lambda $ is chosen as $\lambda \omega +g+\frac{1}{2}\Omega f_{1}(n+1,n)=0$,
namely,
\begin{equation}
(\lambda \omega +g)+\frac{\Omega \lambda }{n+1}e^{-2\lambda
^{2}}L_{n}^{1}(4\lambda ^{2})=0,  \label{PR}
\end{equation}%
the effective Hamiltonian (\ref{ESE}) becomes a Jaynes-Cummings-like model
\begin{equation}
H_{\text{E}}=\omega a^{\dag }a+\omega \lambda ^{2}+2\lambda g+\frac{1}{2}%
\Omega G_{0}(n)\tau _{z}+R_{r}(\tau _{+}a+\tau _{-}a^{\dag }),  \label{JCM}
\end{equation}%
where $R_{r}=2(\lambda \omega +g)$. By means of Hamiltonian (\ref{JCM}), the
analytical energy spectrums and wavefunctions for the Rabi model can be
obtained easily in the subspaces $\{\left\vert +x,n\right\rangle ,\left\vert
-x,n+1\right\rangle \}$ .

\begin{figure}[t]
\includegraphics[width = 0.9\linewidth]{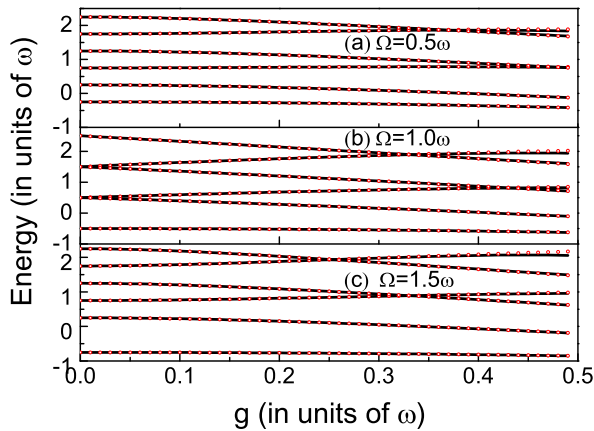}
\caption{(Color online) The energy spectrums of the ground and excited
states for the Rabi model as a function of the interaction strength $g$ for $%
\Omega =0.5\protect\omega $ (a), $\Omega =1.0\protect\omega $ (b), and $%
\Omega =1.5\protect\omega $ (c). The lowest line reflects the ground-state
energy spectrum. In all figures, the black solid lines represent the exact
numerical results, while the red open symbol corresponds to the analytical
results obtained in the paper.}
\label{fig1}
\end{figure}

Before proceeding, the solution of the dimensionless parameter $\lambda $ in
Eq. (\ref{PR}) should be analyzed since it has a crucial role in obtaining
the explicit energy spectrums and wavefunctions. \ In general, the nonlinear
equation (\ref{PR}) cannot be solved analytically. However, in the current
experimental setup with the ultrastrong coupling ($g\leq 0.5\omega $), the
numerical result shows that the dimensionless parameter $\lambda $ is small
compared with the unit. Thus, the associated Laguerre polynomial is given by
$L_{n}^{1}(4\lambda ^{2})\simeq n+1$ since $L_{n}^{1}(y)=n+1+%
\sum_{j>0}c_{j}y^{j}$, and Eq. (\ref{PR}) becomes $(\lambda \omega
+g)+\Omega \lambda e^{-2\lambda ^{2}}=0$, which leads to a solution
\begin{equation}
\lambda \simeq -\frac{g}{\omega +\Omega \exp [-2(\frac{g}{\omega +\Omega }%
)^{2}]}.  \label{LAD}
\end{equation}%
For the weak interaction strength $g$, the dimensionless parameter becomes $%
\lambda \simeq -g/(\omega +\Omega )$.

\begin{figure}[th]
\includegraphics[width = 0.85\linewidth]{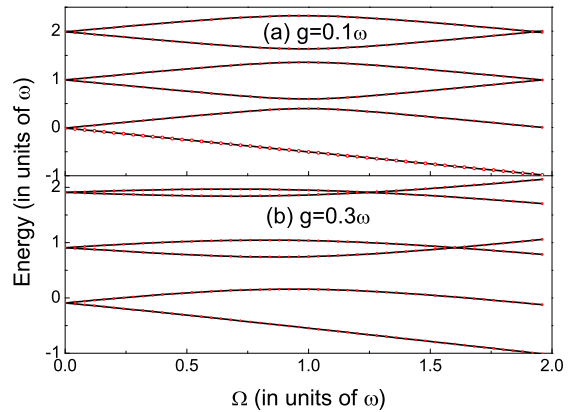}
\caption{(Color online) The energy spectrums of the ground and excited
states for the Rabi model as a function of the resonant frequency $\Omega $
for $g=0.1\protect\omega $ (a) and $g=0.3\protect\omega $ (b). The lowest
line reflects the ground-state energy spectrum. In all figures, the black
solid lines represent the exact numerical results, while the red open symbol
corresponds to the analytical results obtained in the paper.}
\label{fig2}
\end{figure}

\section{Analytical Energy Spectrums}

In terms of the Jaynes-Cummings-like Hamiltonian (\ref{JCM}) and Eq. (\ref%
{LAD}), the ground-state energy spectrum can be written as
\begin{equation}
E_{G}=\omega \lambda ^{2}+2\lambda g-\frac{1}{2}\Omega e^{-2\lambda ^{2}},
\label{GTE}
\end{equation}%
and the excited-state energy spectrum can be given by
\begin{widetext}
\begin{eqnarray}
E_{\pm ,n} &=&(n+\frac{1}{2})\omega +\omega \lambda ^{2}+2\lambda g+\frac{%
\Omega }{4}e^{-2\lambda ^{2}}[L_{n}(4\lambda ^{2})-L_{n+1}(4\lambda ^{2})]
\label{ESEE} \\
&&\pm \frac{1}{2}\sqrt{\{\omega -\frac{\Omega e^{-2\lambda ^{2}}}{2}%
[L_{n}(4\lambda ^{2})+L_{n+1}(4\lambda ^{2})]\}^{2}+4[(\lambda \omega +g)%
\sqrt{n+1}-\frac{\Omega \lambda e^{-2\lambda ^{2}}}{\sqrt{n+1}}%
L_{n}^{1}(4\lambda ^{2})]^{2}} .  \notag
\end{eqnarray}
\end{widetext}

The ground-state energy spectrum in Eq. (\ref{GTE}) is identical to the
result derived from the generalized variational method \cite{Zhang}, which
is, however, invalid for calculating the excited-state energy spectrum. In
addition, for the weak resonant frequency $\Omega $, Eqs. (\ref{GTE}) and (%
\ref{ESEE}) reduce to the results obtained by means of the generalized
rotating-wave approximation \cite{Irish}. In Fig. 1, the energy spectrums of
the ground and excited states for the Rabi model as a function of the
interaction strength $g$ for $\Omega =0.5\omega $ (a), $\Omega =1.0\omega $
(b) and $\Omega =1.5\omega $ (c) are plotted, compared the explicit results
in Eqs. (\ref{GTE}) and (\ref{ESEE}) with the direct numerical simulation.
This figure shows that our analytical energy spectrums including both the
ground- and excited states agree perfectly with the direct numerical
simulation in the current experimental setup with the ultrastrong coupling ($%
g\leq 0.5\omega $). Moreover, these results are valid for all parameter
regimes with the negative detuning $(\Omega <\omega )$, the resonant case $%
(\Omega =\omega )$, and especially the positive detuning $(\Omega >\omega )$%
. This conclusion can be also drawn from Fig. 2, in which the energy
spectrums as a function of the resonant frequency $\Omega $ for $g=0.1\omega
$ (a) and $g=0.3\omega $ (b) are plotted. For the generalized rotating-wave
approximation, the derived energy spectrums can agree well with the
numerical simulation in the case of the negative detuning $(\Omega <\omega )$%
. However, with the increasing of the resonant frequency $\Omega $, this
method breaks down and the error is increased linearly.

\begin{figure}[t]
\includegraphics[width = 0.9\linewidth]{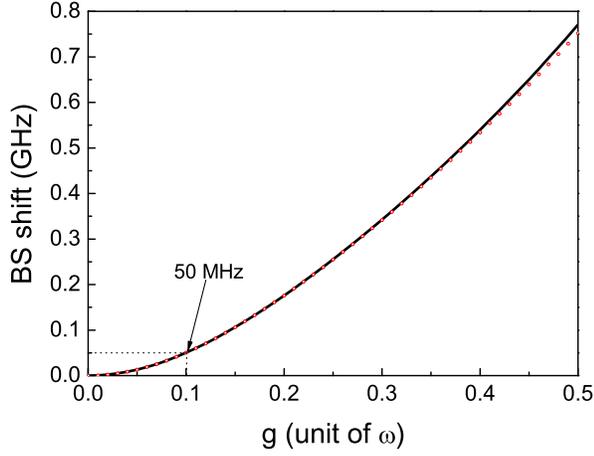}
\caption{(Color online) The Bloch-Siegert (BS) shift as a function of the
interaction strength $g$ with $\protect\omega /2\protect\pi =8.13$ GHz and $%
\Omega /2\protect\pi =4.25$ GHz. In this figure, the black solid lines
represent the exact numerical results, while the red open symbol corresponds
to the analytical results obtained in the paper.}
\label{fig3}
\end{figure}

In addition, by means of Eqs. (\ref{GTE}) and (\ref{ESEE}), the
Bloch-Siegert shift, which is just the energy shift of the level transition,
can be also obtained. In Fig. (3), the smallest Bloch-Siegert shift with the
transition $E_{-,0}\rightarrow E_{G}$ as a function of the interaction
strength $g$ is plotted. The other parameters are the same as those in the
recent experiment of circuit quantum electrodynamics with the ultrastrong
coupling \cite{Fedorov}, namely, $\omega /2\pi =8.13$ GHz and $\Omega /2\pi
=4.25$ GHz. If $g/\omega =0.1$, the smallest Bloch-Siegert shift is $50$
MHz, which agrees well with the experimental observation \cite{Fedorov}.

\section{Analytical Wavefunctions and Mean Photon Number}

The wavefunctions of the Rabi model (\ref{RM}) for the ground and excited
states also need to be discussed. For the ground state, the corresponding
wavefunction can be evaluated immediately as
\begin{equation}
\left\vert \varphi _{0}\right\rangle =e^{\frac{i\pi }{4}\sigma
_{y}}e^{-\lambda \sigma _{z}(a^{\dag }-a)}\left\vert 0,-x\right\rangle .
\label{GSWF}
\end{equation}%
For the excited state, the wavefunction for the Jaynes-Cummings-like
Hamiltonian (\ref{JCM}) is given by
\begin{equation}
\left\{
\begin{array}{c}
\left\vert +,n\right\rangle =\cos \theta _{n}\left\vert n,+x\right\rangle
+\sin \theta _{n}\left\vert n+1,-x\right\rangle \\
\left\vert -,n\right\rangle =-\sin \theta _{n}\left\vert n,+x\right\rangle
+\cos \theta _{n}\left\vert n+1,-x\right\rangle%
\end{array}%
\right. ,  \label{JCWF}
\end{equation}%
where
\begin{equation}
\theta _{n}=\frac{1}{2}\tan ^{-1}\frac{2R_{r}}{E_{-,n}-E_{+,n}},
\label{theta}
\end{equation}%
where $E_{\pm ,n}$ are given by Eq. (\ref{ESEE}). Thus, the excited-state
wavefunction for the Rabi model is given by
\begin{equation}
\left\vert \varphi _{e}\right\rangle =\left\{
\begin{array}{c}
e^{\frac{i\pi }{4}\sigma _{y}}e^{-\lambda \sigma _{z}(a^{\dag
}-a)}\left\vert -,m-1\right\rangle \text{, (}m=1,3,\cdots \text{)} \\
e^{\frac{i\pi }{4}\sigma _{y}}e^{-\lambda \sigma _{z}(a^{\dag
}-a)}\left\vert +,m-1\right\rangle \text{, (}m=2,4,\cdots \text{)}%
\end{array}%
\right. ,  \label{ESWF}
\end{equation}%
where $m$ is the number of the excited state.

\begin{figure}[t]
\includegraphics[width = 0.97\linewidth]{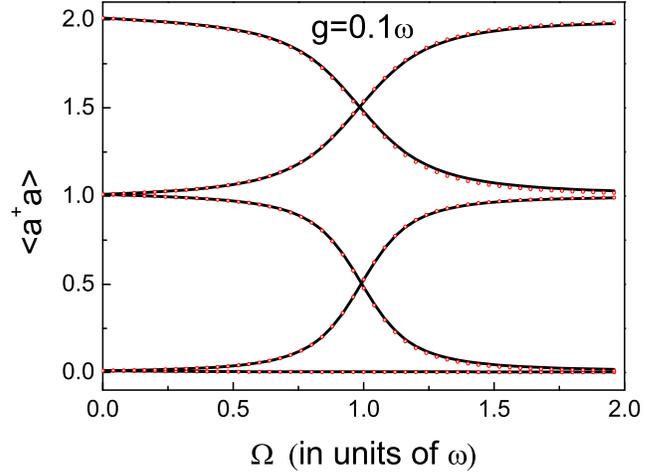}
\caption{(Color online) The mean photon numbers of the ground and excited
states for the Rabi model as a function of the resonant frequency $\Omega $
for $g=0.1\protect\omega $. The lowest line reflects the ground-state mean
photon number. In this figure, the black solid lines represent the exact
numerical results, while the red open symbol corresponds to the analytical
results obtained in the paper.}
\label{fig4}
\end{figure}

Based on the wavefunctions in Eqs. (\ref{GSWF}) and (\ref{ESWF}), the
experimentally measurable physics quantities can be well derived. For
example, the ground-state mean photon number can be given by $\left\langle
a^{\dag }a\right\rangle _{0}=\left\langle \varphi _{0}\right\vert a^{\dag
}a\left\vert \varphi _{0}\right\rangle =$ $\left\langle 0,-x\right\vert
e^{\lambda \sigma _{z}(a^{\dag }-a)}e^{\frac{-i\pi }{4}\sigma _{y}}a^{\dag
}ae^{\frac{i\pi }{4}\sigma _{y}}e^{-\lambda \sigma _{z}(a^{\dag
}-a)}\left\vert 0,-x\right\rangle $, namely,
\begin{equation}
\left\langle a^{\dag }a\right\rangle _{0}=\lambda ^{2}.  \label{MPN}
\end{equation}%
For the excited state, the mean photon number is given by
\begin{equation}
\left\langle a^{\dag }a\right\rangle _{e}=\left\{
\begin{array}{c}
\mu +\lambda ^{2}+\frac{1+2\lambda \tan \theta _{\mu }\sqrt{m+1}}{1+\tan
^{2}\theta _{\mu }}\text{, (}m=1,3,\cdots \text{)} \\
\nu +\lambda ^{2}+\frac{\tan ^{2}\theta _{\nu }-2\lambda \tan \theta _{\nu }%
\sqrt{m+1}}{1+\tan ^{2}\theta _{\nu }}\text{, (}m=2,4,\cdots \text{)}%
\end{array}%
\right. \text{,}  \label{MPET}
\end{equation}%
where $\mu =(m-1)/2$ and $\nu =(m-2)/2$.

In Fig. 3, the mean photon numbers of the ground and excited states for the
Rabi model as a function of the resonant frequency $\Omega $ for $%
g=0.1\omega $ are plotted. This figure shows that the analytical results in
Eqs. (\ref{MPN}) and (\ref{MPET}) agree well with the numerical simulation.
It implies again that the effective Hamiltonian (\ref{JCM}) can describe the
current experimental setup with the ultrastrong coupling. Eq. (\ref{MPN})
also shows that the ground-state mean photon number for the Rabi model
depends on all parameters including the frequency $\omega $ of the quantized
harmonic oscillator, the interaction strength $g$, and especially, the
resonant frequency $\Omega $. It is quite different from the result derived
from the generalized rotating-wave approximation that the ground-state mean
photon number is independent of the resonant frequency $\Omega $ \cite{Irish}%
.

\section{Conclusions}

In summary, we have presented a unitary transformation to map the unsolvable
Rabi model into a solvable Jaynes-Cummings-like model in the dress-state
representation. As a result, the analytical energy spectrums and
wavefunctions including both the ground and the excited states can be
obtained easily. Moreover, our results agree perfectly with the direct
numerical simulations in a wide range of the experimental parameters and are
valid for all regions of the resonant frequency of the two-level system
including the negative detuning $(\Omega <\omega )$, the resonant case $%
(\Omega =\omega )$, and especially the positive detuning $(\Omega >\omega )$%
. Our results can also explain the recent experimental observation in the
circuit quantum electrodynamics with the ultrastrong coupling.

\textbf{Acknowledgments}

We thank Dr. Yuanwei Zhang for his helpful discussion. This work was
supported partly by the 973 program under Grant No. 2012CB921603, the NNSFC
under Grant Nos. 10934004, 60978018, 10904092, 11074154, 11074184 and
61008012, NNSFC Project for Excellent Research Team under Grant No.
61121064, and International Science and Technology Cooperation Program of
China under Grant No.2001DFA12490.

\end{document}